\input susy.sty
\input epsf.sty

\brochureb{\smallsc f. j. yndur\'ain}{\smallsc dynamically generated masses}{1}

\rightline{\petit FTUAM 97-1}
\rightline{\petit UM-TH-97-15}
\rightline{June, 1997\quad} 
\vskip .8cm
\hrule height 0.5mm
\bigskip
\centerline{\titlbf Dynamically Generated Masses in Supersymmetric QCD}
\smallskip 
\centerline{{\titlbf  and Quark Mass Problems}\footnote*{Typeset with \physmatex}}
\medskip
\centerrule{0.8cm}
\bigskip
\bigskip
\fjyuammi
\centerline{\box9}
\bigskip
\bigskip

\setbox0=\vbox{\abstracttype{Abstract}
We consider possible mechanical masses that could appear in supersymmetry, 
other than by direct Higgs coupling to {\sl fermions} and 
 we speculate that the existence of
 such a type of mass would allow one to have the Higgs mass of the 
$u$ quark zero, and the Higgs mass of the $d$ quark 
(at 1 \gev) of $\approx 1\;{\rm to}\;2\;\mev$, 
thus solving at the same time the strong CP problem and arranging the 
grand unification prediction
 $$m_{\mu}/m_{e} = m_{s,\,\rm Higgs}/m_{d,\,\rm Higgs}.$$ 
One possible mechanism for this is  
related to, but not identical with the quark condensate. Here a mass 
is generated which is the same for all quarks, and which adds
 to the Higgs type mass.
 Unfortunately, the numerical value of the generated mas falls short of the desired
 value (some 5~\mev) by orders of magnitude.

An alternate mechanism, through Higgs-induced left-right couplings 
in the squark sector may produce masses of the correct order 
of magnitude if the mixing angles 
are diferent in the  squark and quark sectors. To get 
the desired 
result for the $u$ quark mass, we need 
a stop component mixing of \ffrac{1}{20} for the $LR$ 
$\tilde{u}$ squark  coupling, so the strong 
CP problem may still be solved. For the 
 $d$ mass, this mechanism is not really sufficient to 
solve the grand unification mass ratio problem.}
\centerline{\box0}

\vfill\eject

As is well known, chiral symmetry is dynamically broken in QCD through the nonzero value of
 the quark condensate:
 $$\left< \bar{\psi} \psi\right> =
  \left< {\rm vac}| : \bar {\psi}(0) \psi (0) : |{\rm vac}\right> \neq 0. $$
where $\left. \! | {\rm vac}\right>$ is the physical vacuum.
 This phenomenon occurs even for massless quarks.
The quantity $\left<\bar{\psi} \psi\right>$, while presenting some of
 the properties of a mass term, does
 not imply the generation of a {\it mechanical} mass. This is easily seen by 
noticing
 that insertion of $\left<\bar{\psi} \psi\right>$ in a Green's function (for example,
 the quark propagator)
with momentum $q$ yields contributions behaving like $$q^{-4}\left<\bar{\psi} \psi\right>$$
for large $q$, while a mechanical mass would give $$mq^{-2}.$$
	In this note we remark that, in supersymmetric QCD, broken {\it only} by a mass term
 for squarks and gluinos, 
the quark condensate {\it does} generate a mass. This mass behaves as
 a mechanical mass up to energies
 comparable to the squark, gluino masses, that we take for simplicity
 to be of the same order of magnitude, 
$$m_{\rm gluino}\approx m_{\rm squark} \approx \tilde{m}.$$

An alternate mechanism for the generation of masses for 
quarks occurs if there is a mismatch between the mixing (Cabibbo-Kobayashi-Maskawa) 
angles for quarks and squarks. This would mean that the Higgs-generated 
mass type terms,
$$-\delta_f^2\phi^+_{Lf}\phi^{\phantom {+}}_{Rf}+{\rm h.c.}$$
(the $\phi$ represent squarks, and $f$ is a flavour index) get 
contributions from several flavours. Thus, even if the Higgs-up quark 
coupling would be zero, $\delta_u$ could 
have a nonzero value due to mixing, in particular of squarks  $\tilde{u}$ and $\tilde{t}$. 
One then expects,
$$\delta_u\cong (\sin \tilde{\theta}_{ut})m_t,\;\delta_d\cong (\sin \tilde{\theta}_{db})m_b.$$
We explore possible connections with phenomenological issues.

\brochuresubsection{Dynamically Generated Quark Mass} 
We take our model to be given by the Lagrangian (for simplicity we consider here only one flavor),
$${\cal L}=\bar{\psi_L}\ii\gamma \cdot D \psi_L +
\bar{\psi_R}\ii \gamma \cdot D \psi_R-(1/4)G^2 +
 \bar{\lambda}(\ii \gamma \cdot D-m_{\lambda})\lambda$$
$$+g \sqrt{2}( \phi_L \bar{\psi_L}+\phi_R \bar{\psi_R}) \lambda +{\rm h.c.}
 +\phi_L^+(\dal +\mu^2)\phi_L^{\phantom{+}}+
\phi_R^+(\dal+\mu^2)\phi_R^{\phantom{+}}. \equn{(1)}
$$
Note that as yet we do {\it not} introduce by hand off-diagonal mass terms 
$$\delta^2(\phi_L^+ \phi_R^{\phantom{+}} + \phi_R^+ \phi_L^{\vphantom{+}}),$$
so ${\cal L}$ is invariant under {\it independent} chiral transformations of all 
the {\it L} and {\it R}
fields.

The mechanism for generation of the non-Higgs mechanical mass is simple.
 Consider the left-right piece of the quark propagator, 
$$S = \left<{\rm vac}|{\rm T}\psi_L\bar{\psi}_R | {\rm vac}\right>.$$
\smallskip
\setbox0=\vbox{\hsize 6.truecm \epsfxsize=5.2truecm\epsfbox{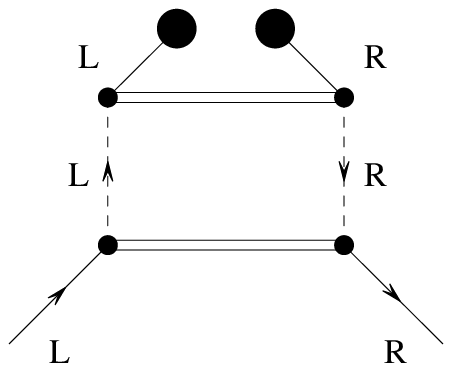}\hfil}
\setbox1=\vbox{\hsize 6.5truecm\captiontype\figurasc{figure 1.}{Diagram yielding mass-like term 
from the quark condensate.}\hb
\vskip.4cm}
\line{\quad\tightboxit{\box0}\hfil\box1\quad}
\smallskip
The relevant effect is generated by the diagram of \fig~1. The large blobs there represent the
 vacuum expectation value of the quark condensate. The corresponding contribution to $S$, $S_m$ 
is evaluated with standard techniques. One gets the result
$$\eqalign{S_m =(-1/3p^2)\ii g^4 (1+\gamma_5)(C_F^2/N_c) m^2_{\lambda}\cr
\times \int {\dd^4k \over (2 \pi)^4}{1 \over {(p-k)^2-m^2}}  {1 \over {(k^2- \mu^2)^2}}
{1 \over {k^2 -m^2_{\lambda}}} \left<\bar {\psi} \psi\right>\cr
\approx i{- \alpha_s^2 C_F^2 \over 3p^2N_c} (1+ \gamma_5){\left<\bar{\psi}
\psi\right> \over 6 \tilde m^2},}
\equn{(2)}$$
the last expression for 
$|p^2| \ll m_{\lambda} \approx \mu \approx \tilde m.$
This corresponds, for momenta $q^2$ such that $\Lambda^2 \ll |q^2| \ll \tilde m^2$
 (with $\Lambda$ the QCD parameter) to a generated mechanical mass 
$$m_{\bar{\psi}\psi} = {2C_F^2 \alpha_s^2 \over 9N_c}\, {-\left<\bar{\psi}\psi \right> \over \tilde m^2}.
 \equn{(3)} $$

\brochuresubsection{Applications} Eqs. (2), (3) have been obtained for quarks with, initially, zero
 mechanical mass. If we take a quark with non-zero mass $m_{\rm Higgs}$ generated by the Higgs
 mechanism, with $m_{\rm Higgs} \ll \Lambda$,
then we get a total mass
$$m=m_{\rm Higgs}+ m_{\bar{\psi}\psi}.$$
The existence of a mass generated by a mechanism other than the Higgs one would
be welcome for at least two reasons. To discuss them,
 we first remark that all experimentally based
 determinations of the masses of the light quarks \ref{1,2} yield the {\it total}
mass, $m$. From the standard values of the $\rm {\overline{MS}}$-renormalized masses,
 at the reference momentum of 1~\gev, we then have
$$m_u = 5\;\mev, m_d = 8\;\mev, m_s = 200\;\mev. \equn{(4)}$$
Now, we can assume that {\it all} the {\it u} quark mass is {\it generated dynamically}\fnote{
From the existence of {\sl lower} bounds on quark masses, 
 independent on assumptions about experimentally inaccessible 
pieces of the pseudoscalar correlator (ref. 2) it follows 
that the possibility $m_u=0$ cannot be contemplated 
seriously as a solution to the strong CP problem.}.
 Thus, the Higgs mass of this quark would vanish (apparently, this occurs 
naturally in a class of supersymmetric models\ref{3})
 and we would preserve the two independent chiral
invariances  
${\rm U}(1)_L \times {\rm U}(1)_R$, which is sufficient to solve the strong CP
problem. This assumption on the mass of the $u$ quark then implies
$$m_{\bar{\psi}\psi} = 5\;\mev$$
and therefore we get the second bonus: from this and the values of the masses in 
 Eq.(4) it follows that
$$m_{d,\,{\rm Higgs}} = 3 \;\mev,$$
which goes in the righ direction towards solving the outstanding mass puzzle\ref{4} that occurs in 
grand unified SU(5), with minimal Higgs system. It will be remembered that there one has
$$m_{b,\,{\rm Higgs}} = m_{\tau},$$
a relation well satisfied if taking into account renormalization effects. However,
 the same assumptions give
$$m_s/m_d=m_{\mu}/m_e. \equn{(5)} $$
This relation is independent of renormalization effects, and is violated by a
 full order of magnitude if we take the masses in (5) to mean the full masses: note that 
\equn{(4)} gives
 $$m_s/m_d=25.$$
 However, relation (5) actually
 follows for the {\it Higgs} masses, and then
$$m_{s,\,{\rm Higgs}}/m_{d,\,{\rm Higgs}}=67,$$
and by fiddling with the errors of (4) one can even get to the experimental $\mu /\rm e$ 
ratio.

\brochuresubsection{Discussion} The previous analysis has made it clear that it would be
 very desirable to have, beyond the Higgs mass, a dynamically generated mass of the order of
$m_u$. The mechanism discussed at the begining of this note provides such a mass.
Unfortunately, it is far too small. If we assume ${\tilde m} \approx 100\;\gev$,
 and renormalize at 100 GeV, then we get
$$m_{\bar {\psi} \psi} \approx 2.1 \times 10^{-6}\;\mev,$$
way off the necessary $$m_u(100\;\gev) \approx 2.7\;\mev.$$

It is of course possible to get the desired effects by introducing, in an {\it ad hoc} manner,
a condensate
 $$\left< \phi_L^+ \phi_R^{\phantom{+}} \right> \not= 0, $$
or, equivalently, by assuming a coupling
$$-\delta^2\phi^+_L\phi_R^{\phantom{+}}+{\rm h.c.}.\equn{(6})$$
Let us consider the second for definiteness, and because it 
arises somewhat naturally in the standard supersymmetric context. 
Assume again degenerate SUSY masses 
$\tilde{m}$ for 
simplicity. The term (6) still respects the two independent ${\rm U}(1)_L \times {\rm U}(1)_R$, 
{\it for fermions\/}:  this is sufficient for our aims. The mass generated, given 
by the diagram of \fig~2, is easily computed, 
and we get, for momenta much smaller than $\tilde{m}$, 
$$m_{\phi^+_L\phi_R^{\phantom{+}}}=\dfrac{\delta^2\alpha_s(\tilde{m}^2)}{4\pi\tilde{m}}.
\equn{(7)}$$
\setbox0=\vbox{\hsize 6.5truecm \epsfxsize=6.4truecm\epsfbox{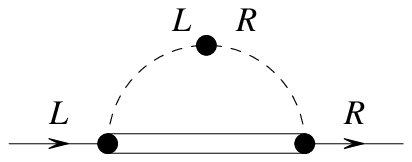}}
\setbox1=\vbox{\hsize 6.truecm\captiontype\figurasc{figure 2.}{Diagram yielding mass term 
from $LR$ squark coupling.}\hb
\vskip.4cm}
\smallskip
\line{\quad\tightboxit{\box0}\hfil\box1\quad}
\smallskip
For $\tilde{m}=200\gev$, agreement with the phenomenological 
value of the $u$ quark mass, still with with zero Higgs mass, is obtained
 if $\sqrt{\delta_{u}^2}\simeq 8 \gev$, \lie,
 $\sin \tilde{\theta}_{ut}\sim 1/20$.
  
This mechanism therefore appears to give a possible solution to the strong CP problem, 
although it looks a bit contrived: 
there seems to be no apparent reason why the angle $\tilde{\theta}_{ut}$ should 
be so much larger than the corresponding quark mixing angle, $O(10^{-3})$. The  
 problem of the mass ratio, \equn{5}, is however left unsolved.  To get 
a contribution to 
the phenomenological value of the $d$ quark mass so that its  Higgs 
mass is left with only $\sim 2\,\mev$ one also needs $\sqrt{\delta_{d}^2}\simeq 8 \gev$, 
which is impossible as this is larger than the bottom quark mass. One could try to 
repair this with a mixed mechanism: it is possible to imagine that the {\sl electron} 
observed mass is now the {\sl difference} between a larger Higgs 
mass, and a SUSY generated mass, due to mixing 
of the Higgs couplings of $\tilde{e}$ and $\tilde{\tau}$, 
so one would have $m_e=m_{e,\,{\rm Higgs}}-m_{e,\, \phi^+\phi}$. Renormalizing the masses 
at the SUSY scale, $\tilde{m}\simeq 200\,\gev$, we have
$$\eqalign{m_b\simeq 3\,\gev,\;m_{\tau}\simeq 2\,\gev,\,\alpha_s\simeq 0.1\cr
{\rm and\qquad}\cr
m_e=0.5\,\mev,\,m_d=4.5\,\mev,\,m_s=150\,\mev\cr}$$
so that use of \equn{(7)} for the Higgs components 
gives
$$\dfrac{m_{d,\,{\rm Higgs}}}{m_{e,\,{\rm Higss}}},
\;m_{\phi^+\phi}\simeq 2.5\,\mev,$$
which is hardly possible, even with mixing angles of
 $\tilde{\delta}_{b}\simeq\tilde{\delta}_{\tau}\simeq \pi/2$ \ --quite an 
extreme situation indeed.      

 In summary, we may say that the mechanism that is natural does
 not work phenomenologically, and the one that works phenomenologically 
lacks sufficient theoretical justification. The results may be relevant 
to the strong CP problem, but less likely so for the unification 
of fermion masses: certainly not unless unexpected physics enter the game.
\vfill\eject
\brochuresubsection{Acknowledgements}
Discussions with G. Kane on possible supersymmetric scenarios are also gratefully 
ackowledged.
\brochuresubsection{References}
\item{1.}\ajg{C87}{1982}{77}{J. Gasser and H. Leutwyler}{Phys. Rep.};
 \ajg{174}{1987}{372}{C. A. Dom\'\i nguez and E. de Rafael}{Ann. Phys. {(\rm N.Y.)}}. 
\item{2.}{\sc C. Becchi et al.,}\jzp{C8}{1981}{375}
\item{3.}See, \leg, {\sc P. Binetruy, S. Laynac and P. Ramond}, \jnp{B477}{1996}{353},
 and work quoted there.
\item{4.}\ajg{1}{1979}{167}{F. J. Yndur\'ain}{Kinam}.
\vfill\eject

\end